\documentclass[pra,showpacs,twocolumn,superscriptaddress]{revtex4}

\usepackage[utf8]{inputenc}
\usepackage[T1]{fontenc}

\usepackage{graphicx}
\usepackage{dcolumn}
\usepackage{bm}
\usepackage{amsmath}
\usepackage{color}
\usepackage{gchords}

\usepackage{ulem}

\newcommand{\bra}[1]{\left\langle #1 \right|}
\newcommand{\ket}[1]{\left| #1 \right\rangle}

\newcommand{\ketbra}[2]{\left|#1\middle\rangle\middle\langle#2\right|}

\newcommand{\mean}[1]{\left\langle#1\right\rangle}

\newcommand{\M}[1]{\mathcal{#1}}

\newcommand{\ad}[1]{a^{\dagger}_{#1}}

\begin{document}
\title{Entanglement of indistinguishable particles as a probe for quantum phase transitions
in the extended Hubbard model}

\author{Fernando Iemini}
\email{fernandoiemini@gmail.com}
\author{Thiago O. Maciel}
\author{Reinaldo O. Vianna}
\affiliation{Departamento de F\'{\i}sica - ICEx - Universidade Federal de Minas Gerais,
Av. Pres Ant\^onio Carlos 6627 - Belo Horizonte - MG - Brazil - 31270-901.}

\date{\today}

\begin{abstract} 
We investigate the quantum phase transitions of the extended Hubbard model at half-filling with
periodic boundary conditions employing the entanglement of particles, as opposed to the more traditional
entanglement of modes.
  Our results show 
 that the entanglement  has either discontinuities or local minima
  at the critical points.  We associate the
   discontinuities to first order transitions, and the minima to 
   second order ones. Thus we show that the entanglement of particles 
   can be used to derive the phase diagram, except for the subtle transitions between
   the phases SDW-BOW,  and the superconductor phases TS-SS.  
\end{abstract}

\maketitle

\section{Introduction}

The connection between two
 important disciplines of Physics, namely
 quantum information theory and
 condensed matter physics, has been the subject of great interest recently,
generating much  activity
 at the border of these fields, with numerous interesting 
questions addressed so far \cite{amico08}. In particular, 
  the properties
 of  entanglement in many-body systems, and 
 the analysis of its  behavior  in critical systems deserve special attention.
   
In this work we deal with the
 entanglement of indistinguishable fermionic particles
 in the one dimensional extended Hubbard model (EHM). 
 We focus in
  the half-filling case.
 The model is a generalisation of the Hubbard model
 \cite{hubbard63, solymon}, which 
 encompasses more general interactions between the
 fermionic particles, such as an inter-site interaction, thus describing
 more general phenomena and a richer phase diagram. Precisely, it is given by,
\begin{eqnarray}
H_{EHM}&=&-t\sum\limits_{j=1}^{L}\sum\limits_{\sigma = \uparrow, \downarrow}
(\ad{j,\sigma}a_{j+1,\sigma} + \ad{j+1,\sigma}a_{j,\sigma}) +\nonumber \\
& & +\, U\, \sum_{j=1}^{L}\hat{n}_{j\uparrow}\hat{n}_{j\downarrow}
 + V\, \sum_{j=1}^{L}\hat{n}_{j}\hat{n}_{j+1},
\label{ehm}
\end{eqnarray}
where $L $ is the lattice size, $\ad{j,\sigma}$ and $a_{j,\sigma}$ are
 creation and annihilation operators, respectively, of a fermion with spin
 $\sigma$ at  site $j$,
 $\hat{n}_{j,\sigma} = \ad{j,\sigma}a_{j,\sigma}$, $\hat{n}_{j}
 = \hat{n}_{j,\uparrow} + \hat{n}_{j,\downarrow}$, and we consider
 periodic boundary conditions (PBC), $L+1=1$. The hopping
 (tunnelling) between neighbor sites is parametrized by $t$,
 while the on-site and inter-site interactions are given by
 $U$ and $V$, respectively. Despite the apparent simplicity
 of the model, it exhibits a very rich phase diagram, which
 includes several distinct phases, namely: charge-density wave (CDW),
 spin-density wave (SDW), phase separation (PS), singlet (SS)
 and triplet (TS) superconductors, and a controversial bond-order
 wave (BOW). A more detailed description of the model and
 its phases will be given in the next section.

Our numerical analysis is performed employing entanglement measures
 for indistinguishable particles introduced recently \cite{iemini13a, iemini13b,iemini14},
in conjunction with the density-matrix renormalisation
 group approach (DRMG)\cite{DMRG, ALPS}, 
which has established itself as a
 leading method for the simulation of one dimensional strongly
 correlated quantum lattice systems. DMRG is a numerical
 algorithm for the efficient truncation of the Hilbert space
 of strongly correlated quantum systems based on a rather
 general decimation prescription. The algorithm has achieved 
unprecedented precision in the description of static, dynamic
 and thermodynamic properties of one dimensional quantum systems, 
quickly becoming the method of choice for numerical studies.

The paper is organised as follows. In Sec. \ref{ext.hub.model} we review 
the model and its phase diagram. In Sec. \ref{ent.ind.part} we present the distinct 
definitions of entanglement in systems of indistinguishable particles, focusing on the
 notion of ``entanglement of particles''.
  In Sec. \ref{ent.and.qpt} we present our results. We conclude in Sec. \ref{conclusion}.  

\section{Extended Hubbard Model}
\label{ext.hub.model}
 
 In this section we give a detailed description of the extended
 Hubbard model \cite{hubbard63, solymon}, and its distinct phases in the half-filling case.
   The reader familiar with
 the subject may skip this  section.
  
\begin{figure}
\centering
\includegraphics[scale=0.32]{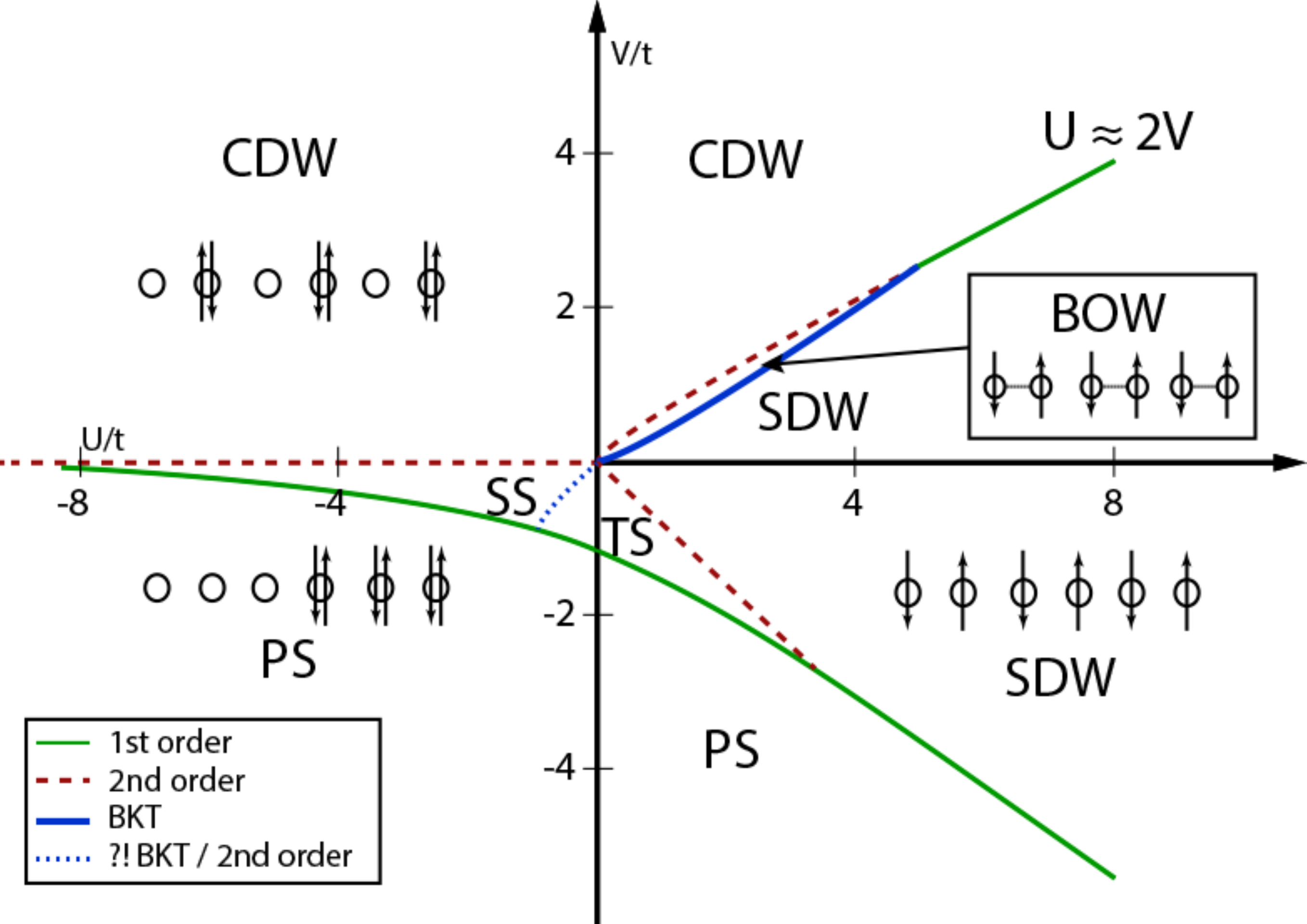}
\caption{ (Color online) Phase diagram of the half-filled extended Hubbard
 model in one dimension. The distinct phases correspond
 to the charge-density wave (CDW), the spin-density wave (SDW), phase
 separation (PS), singlet (SS) and triplet (TS) superconducting phases,
 and bond-order wave (BOW). 
  The order of the quantum phase transitions is identified 
 by the different line shapes. The order of the two superconducting 
 phases transition (blue dotted line) is controversial, being identified 
 as a BKT transition \cite{nakamura00}, or a 
 second order transition \cite{lin00}.}
\label{phase.diagram.ehm}
\end{figure}

Many efforts have been devoted to the investigation of the EHM's
 phase diagram at half-filling, using both analytical
 and numerical methods
 \cite{lin00,nakamura00,sengupta02,zhang04,dalmonte14,jeckelmann02,ejima07,sandvik04}.
 Despite the apparent simplicity of the model, it exhibits a
 very rich phase diagram which includes several distinct phases:
 charge-density wave (CDW), spin-density wave (SDW), phase
 separation (PS), singlet (SS) and triplet (TS) superconductors,
 and a controversial bond-order wave (BOW). See
 Fig.\ref{phase.diagram.ehm} for a schematic drawing of
 the phase diagram at half-filling.

In the strong coupling limit ($|U|,|V|\gg t$), one can qualitatively
 characterize its phases as  given by a charge-density wave,
 spin density wave and a phase separation. For a strong 
repulsive on-site interaction ($U>0$, $U \gg V$), the ground 
state avoids double occupancy  and the spin density
 is periodic along the lattice, leading to an antiferromagnetic 
ordering, namely spin-density wave.
Its order parameter is given by,
\begin{equation}
\mathcal{O}_{sdw}(k) = \frac{1}{L}\sum_{m,n} e^{ik(m-n)}
 \left[ \mean{\sigma^z_m \sigma^z_n} -  \mean{\sigma^z_m}
 \mean{\sigma^z_n} \right],
\end{equation}
where $\sigma_j^z = \frac{1}{2}(\hat{n}_{j \uparrow}-\hat{n}_{j 
\downarrow})$. In the limit $U \rightarrow \infty$, the ground
 state is dominated by the following configurations:
\begin{eqnarray}
\ket{\psi}_{sdw} \approx \frac{1}{\sqrt{2}} \left( \,\ket{ 
\uparrow, \downarrow,\uparrow, \downarrow, ...\, , \uparrow_{L-1},
 \downarrow_{L}  } + \right. \nonumber \\
+  \left. \ket{ \downarrow,\uparrow, \downarrow,\uparrow, 
 ...\, , \downarrow_{L-1}, \uparrow_{L} } \, \right),
  \label{sdw.phase.stronglimit}
\end{eqnarray}
where the state is described in the real space mode representation, 
in which each site
 can be in the following set of configurations: $\{\ket{0},\ket{\uparrow},\ket{\downarrow},
 \ket{\uparrow \downarrow}\}$. 
Considering a strong repulsive inter-site interaction ($V>0$, 
$V \gg U$), a periodic fermionic density is generated,
 leading to a charge-density wave. 
 Its order parameter is given by,
\begin{equation}
\mathcal{O}_{cdw}(k) = \frac{1}{L} \sum_{m,n} e^{ik(m-n)}
 \left[ \mean{\hat{n}_m \hat{n}_n} -  \mean{\hat{n}_m} 
\mean{\hat{n}_n} \right].
\end{equation}
In the limit $V \rightarrow \infty$, the ground state is
 dominated by the following configurations,
\begin{eqnarray}
\ket{\psi}_{cdw} \approx \frac{1}{\sqrt{2}}\,  \left(
 \,\ket{ \uparrow \downarrow,0, \uparrow \downarrow,
 0 , ...\, , \uparrow \downarrow _{L-1}, 0 } + \right.
 \nonumber \\
+  \left.  \ket{ 0, \uparrow \downarrow, 0, \uparrow 
\downarrow, ...\, ,0, \uparrow \downarrow _{L} }\,\right). 
 \label{cdw.phase.stronglimit}
\end{eqnarray}
In the range of strong attractive interactions ($U,V<0$
or $U>0$, $V<0$ with $|V|\gg |U|$), the fermions cluster
 together, and the ground state becomes inhomogeneous, with
 different average charge densities in its distinct spatial
 regions. Such a phase is called phase separated state. In 
the limit $V \rightarrow -\infty$, the ground state is
 dominated by the following configurations,
\begin{eqnarray}
\ket{\psi}_{ps} &\approx &\frac{1}{\sqrt{L}}
 \,\sum_{\{\hat{\Pi}\}} \hat{\Pi}  \ket{ \uparrow \downarrow,
 \uparrow \downarrow,  ...\, , \uparrow \downarrow_{(\frac{L}{2})}, 0, ...\, , 0},
  \label{ps.phase.stronglimit}
\end{eqnarray}
where $\{\hat{\Pi}\}$ is the set of translation operators.

In the weak coupling limit, different phases appear. For
 small attractive inter-site interactions ($V<0$),  
superconducting phases are raised, characterized by the
 pairing correlations,
\begin{eqnarray}
\Delta_x = \frac{1}{\sqrt{L}} \sum_j a_{j,\uparrow} a_{j+x,\downarrow},
\end{eqnarray}
with the respective order parameter $\mathcal{O}_s  =
 \sum_{x,x'} \mean{\Delta_x^{\dagger} \Delta_{x'} }$. If 
the on-site interactions are lower than the inter-site 
interactions ($U\leq 2V$),  the fermions will pair as 
a singlet superconductor, characterized by  nearest-neighbor
 ($\Delta_{ss_{nn}}$) or on-site ($\Delta_{ss_{o}}$) singlet 
pairing correlations given by, 
\begin{eqnarray}
\Delta_{ss_{nn}} &=& \Delta_x - \Delta_{-x} \nonumber \\
&=&  \frac{1}{\sqrt{L}} \sum_j (a_{j,\uparrow} a_{j+x,\downarrow}
 - a_{j,\downarrow} a_{j+x,\uparrow}), \\ 
\Delta_{ss_{o}} &=& \Delta_0  =  \frac{1}{\sqrt{L}} \sum_j 
a_{j,\uparrow} a_{j,\downarrow}, 
\end{eqnarray}
where $x=1$. On the other hand, if the on-site interactions
 are higher than the inter-site interactions ($U\geq2V$),
  we have a triplet superconductor, characterized by 
nearest-neighbor triplet pairing correlations ($\Delta_{ts_{nn}}$)  given by,
\begin{eqnarray}
\Delta_{ts_{nn}} &=& \Delta_x + \Delta_{-x} \nonumber \\
&=&  \frac{1}{\sqrt{L}} \sum_j (a_{j,\uparrow} a_{j+x,\downarrow} 
+ a_{j,\downarrow} a_{j+x,\uparrow}),
\end{eqnarray}
where $x=1$.

Note that the difference between the
 singlet and triplet pairing correlations is simply a plus 
or minus sign. It can be clarified if we consider,
 for example, the case of two fermions in a singlet or 
triplet spin state, given  by $\left( \ket{ij}
 \pm \ket{ji} \right)\,\left( \ket{\uparrow \downarrow} \mp
 \ket{\downarrow \uparrow} \right)$. Expanding this state, we have,
\begin{eqnarray}
& &\ket{i j} \left( \ket{\uparrow \downarrow} \mp 
\ket{\downarrow \uparrow} \right) \pm \ket{ji}
 \left( \ket{\uparrow \downarrow} \mp \ket{\downarrow \uparrow}
 \right) \nonumber \\ 
&=& \ket{i \uparrow, j \downarrow} \mp \ket{i \downarrow, j
 \uparrow}  \pm \ket{j \uparrow, i \downarrow}  - 
\ket{j \downarrow, i \uparrow} \nonumber \\
&=& \left(\ket{i \uparrow, j \downarrow} - \ket{j \downarrow,
 i \uparrow} \right) \mp \left( \ket{i \downarrow, j \uparrow}
  - \ket{j \uparrow, i \downarrow} \right) \nonumber \\
&=& \left(\ad{i \uparrow}\ad{j \downarrow} \mp \ad{i \downarrow}
\ad{j \uparrow}\right) \ket{vac},
\end{eqnarray}
where the  {lower/upper} sign corresponds to the
 {singlet/triplet} pairing correlation.

 The last phase in the diagram is the controversial
 bond-order-wave (BOW). By studying the EHM ground state
 broken symmetries, using level crossings in excitation 
spectra, obtained by exact diagonalization, Nakamura 
\cite{nakamura00} argued for the existence of a novel 
bond-order-wave  phase for small to intermediate values
 of positive U and V, in a narrow strip between  CDW and
  SDW phases. This phase exhibits  alternating
 strengths of the expectation value of the kinetic energy
 operator on the bonds, and is characterized by the following order parameter,
 \begin{eqnarray}
 \mathcal{O}_{bow}(k) = \frac{1}{L}\sum_{m,n} e^{ik(m-n)}
 \left[ \mean{B_{m,m+1} B_{n,n+1}} \right. \nonumber \\ 
\left. - \mean{ B_{m,m+1}} \mean{ B_{n,n+1}  } \right],
 \end{eqnarray}
 where 
$B_{m,m+1} = \sum_{\sigma}(\ad{m,\sigma}a_{m+1,\sigma} + H.c.)$
   is the
 kinetic energy operator associated with the $m$th bond.
  Nakamura argued that the CDW-SDW 
transition is replaced by two separate transitions, namely: (i) a 
continuous CDW-BOW transition; and (ii) a
 Berezinskii-Kosterlitz-Thouless (BKT) spin-gap transition 
from BOW to SDW. Such  remarkable proposal was later 
confirmed by several works
 \cite{sengupta02,zhang04,dalmonte14,jeckelmann02,ejima07,sandvik04},
 employing different numerical methods, like
 DMRG, Monte Carlo
 or exact diagonalization. Nevertheless, while the BOW-CDW phase
 boundary can be  well resolved, since it involves a 
standard second order (continuous) phase transition, the
 SDW-BOW boundary is more difficult to locate, for it involves 
a BKT transition in which the spin gap opens exponentially
 slowly as one enters the BOW phase. The precise location 
of the BOW phase is then still a subject of debate. To the best of our knowledge, the best
 estimates for the transitions, taking  $U/t=4$, 
correspond to a CDW-BOW transition 
at $V/t\approx 2.16$ \cite{sengupta02,zhang04,ejima07,sandvik04}, 
and to a BOW-SDW transition in 
the range $V/t \approx 1.88-2.00$ \cite{sengupta02,zhang04,ejima07,sandvik04},  
or $V/t= 2.08\pm0.02$ \cite{dalmonte14}.

\section{Entanglement of indistinguishable particles}
\label{ent.ind.part}

Despite widely studied in systems of
 distinguishable particles, entanglement or  more general notions of quantum correlations have received  less attention
 in the case of indistinguishable particles.
  In this case, 
the space of quantum
states is restricted to symmetric ($\M{S}$) or antisymmetric ($\M{A}$)
subspaces, depending on the bosonic or fermionic nature of the
system, and the particles are no longer accessible individually,
 thus eliminating the usual notions of separability 
and local measurements, and
 making the analysis of correlations much subtler.
In fact, there are a multitude of distinct approaches and
 an ongoing debate around the entanglement
 in these systems \cite{balachandran,ghirardi02,ghirardi04,iemini14,
schliemann01a01b,eckert02,li01,barnum04,somma04,zanardi02,wiseman03,benatti10,banulus07,
grabowski1112,tsutsui}. Nevertheless, despite the variety,
 the approaches consist essentially in the
analysis of correlations under two different aspects: the
correlations genuinely arising from the entanglement between
the particles (entanglement of particles) \cite{balachandran,ghirardi02,ghirardi04,iemini14,
schliemann01a01b,eckert02,li01,barnum04,somma04},
 and the correlations arising from the
entanglement between the modes of the system (entanglement of modes)
 \cite{zanardi02,wiseman03,benatti10,banulus07}.
 These two notions of entanglement are
complementary, and the use of one or the other depends
on the particular situation under scrutiny. For example,
the correlations in eigenstates of a many-body Hamiltonian
could be more naturally described by  entanglement of particles,
whereas certain quantum information protocols could prompt
a description in terms of entanglement of modes.
 Once one has opted for a certain notion of entanglement,
  there are interesting methods  to quantify it
\cite{iemini13a, iemini13b, iemini14, paskauskas01,plastino09,zander10,reusch15}.

Entanglement of modes can be understood by mapping the quantum state in its number representation, namely,
\begin{eqnarray}
\ad{j_1 } ... \ad{j_N } \ket{vac} &\longrightarrow &
\ket{0 ... 1_{j_1 } ... 1_{j_N } ... 0}\nonumber \\ 
\hat{\mathcal{A}}(\mathcal{H}_1^{M} \otimes ... \otimes \mathcal{H}_N^{M}) &\longrightarrow &
  (\mathcal{H}_1^{2} \otimes ... \otimes \mathcal{H}_{M}^{2})
\end{eqnarray}
where {$j_i = 1,...\,,M$, and} $\{\ad{j}\}_{j=1}^M$ is an arbitrary set of $M$ fermionic operators  describing 
the single particle modes of the system {(not necessarily 
the real space modes as in the Hamiltonian definition)}. 
We will denote hereafter as ``configuration 
representation (number representation)'' the left (right) side of 
the previous equation. Such equation corresponds to a mapping to 
distinguishable qubits, represented by the occupied ($\ket{1}_{j}$ )
 or unoccupied ($\ket{0}_{j}$) {modes}, which then allows 
 one to employ all the tools commonly
 used in distinguishable quantum systems in order to analyze their 
 correlations. One could, for example, use the von Neumann entropy 
 of the reduced state representing a block with $\ell$ 
  {modes}, in 
 order to quantify the entanglement between this block with the rest 
 of the {modes}. The reduced state is obtained by the partial 
 trace in the number representation  ($ \rho_{\ell} = Tr_{j \notin 
 \ell} \left(\ketbra{\psi}{\psi}\right)$). Notice that, in 
 the mode representation, {\it local}  observables  may actually 
 involve correlations between  particles. For example,  in the Hubbard model,  although the operator 
 ``$\ad{j\uparrow}
 \ad{j\downarrow} a_{j\downarrow} a_{j\uparrow}$''
 acts locally 
 at the $j$th site {(real space modes)}, 
 it describes pairing correlations between  particles. The 
 algebra of local observables at the {modes}, defined in the number representation,
 is generated by,
 \begin{eqnarray}
 \Omega_{loc} = \left\{ \hat O_{1} \otimes \mathcal{I}_{2,M} \,;\,
 \mathcal{I} \otimes \hat O_{2} \otimes 
 \mathcal{I}_{3,M}\,;\, \cdots \, \nonumber \right. \\
\left. \cdots \,;\, \mathcal{I}_{1,(M-1)} \otimes 
\hat O_{M} \,\,||\,\, \hat O_{j}^{\dagger} = 
\hat O_{j} \right\}
 \end{eqnarray}
 where $\mathcal{I}_{i,j} \equiv \mathcal{I}_i \otimes 
 \mathcal{I}_{i+1} \otimes \cdots \otimes \mathcal{I}_j$, with $j>i$ and 
 $\mathcal{I}_i$ is the identity operator acting on mode $i$. In this way, unentangled 
 states are those which can be completely described 
 by such local observables. It is known that such 
 states are simply the separable states in the usual tensor 
 product form, $\ket{\psi}_{un} = \ket{\phi_{1}} 
 \otimes \ket{\phi_{2}}  \cdots \otimes 
 \ket{\phi_{M}}$.

Based on  the previous reasoning, we now define the notion 
 of {\it entanglement of particles}. 
 Notice first that one cannot analyze the system under the usual
 paradigm of  {\it separability and locality}, where  the reduced states obtained by partial trace are mixed
 ($\rho_r  = Tr_{2...N}(\ketbra{\psi}{\psi})$), whenever the global state is pure and entangled.
Therefore, in the case of indistinguishable particles in the configuration representation,  
  the use of partial trace
   to characterize entanglement
   should be carefully reviewed, since it would suggest that all  pure fermionic states
  are  entangled, given that their reduced states are always mixed.
 In order to   generalize the notion 
 of entanglement for systems of indistinguishable particles,  
 the approach based on the algebra of observables 
 sheds light on the problem  and allows us to go beyond the
 paradigm of separability and locality. 
 
 We now define the proper algebra of ``local observables'' as the 
 one composed by operators which do not create correlations 
 between the indistinguishable particles. Such algebra, defined 
 in the configuration representation, is generated by the 
following  single particle operators,
\begin{eqnarray}
\Omega_{loc} = \left\{ \hat O \otimes \mathcal{I}_{2,N} \,+\, \mathcal{I} \otimes \hat O \otimes \mathcal{I}_{3,N}
 \,+\, \cdots \right. \nonumber \\
 \left. \cdots \,+\, \mathcal{I}_{1,(N-1)} \otimes \hat O \,\, || \,\, \hat O^{\dagger} = \hat O \right\},
\end{eqnarray}
where $N$ is the number of particles. Equivalently, 
using the second quantization formalism, the above set is given 
by the number conserving quadratic operators, $\Omega_{loc} 
= \{ (\ad{i} a_j + H.c.) \,|\, i,j=1,...,M\}$. 
The states that can be completely described by such algebra form, in this way, the set of unentangled states,
 where any particle is not entangled with any other. Intuitively, we would expect that this set corresponded to single Slater determinants with fixed particle number.
 More precisely, for a system with $N$ fermions, it is given by,
\begin{equation}
\ket{\psi}_{un} = a^{\dagger}_{j_1} 
a^{\dagger}_{j_2} ... a^{\dagger}_{j_N} \ket{vac},
\label{unentangled.pure.state}
\end{equation}
where $\{ a^{\dagger}_{j} \}$ is an arbitrary 
set of fermionic operators. Recall  that these operators 
  cannot be quasiparticles with particle-hole superpositions, 
  as usual in a Bogoliubov transformation, since the above 
   states have a fixed number of fermions.
 In fact, 
  distinct approaches confirmed that
   such set does indeed correspond to the  unentangled states 
    \cite{balachandran,barnum04,somma04,ghirardi02,ghirardi04,iemini14,
schliemann01a01b,eckert02,li01}.  The only 
  non-classical correlation present in such states is the {\it exchange},
 due to the antissymetrization, which does not constitute 
 entanglement. For example, in \cite{balachandran} the analysis 
 follows by using a very elegant mathematical formalism, called GNS 
  (Gelfand-Naimark-Segal) construction, for the case of
 two fermions, each one with Hilbert space dimension 3 or 4, and two bosons with dimension 
 3; in \cite{barnum04,somma04}
 the authors propose a ``Generalized Entanglement (GE)'' measure, obtaining
  a simple formula for the ``partial trace'',
  and the set of fermionic unentangled states
 for an arbitrary number of particles; or also in \cite{iemini14},
where a general notion
 of quantum correlation beyond entanglement (the \textit{quantumness} of correlations)
 is investigated by means of an
  ``activation protocol'', which yields
 the same set
 of states with no \textit{quantumness} as the above unentangled one.

As in the case of distinguishable modes, the von Neumann entropy
 also provides a good quantifier 
for the entanglement of indistinguishable particles. We can define
 the Shifted von Neumann entropy of 
entanglement \cite{iemini13b} as follows,
 \begin{equation}
E_p(\ketbra{\psi}{\psi}) = S(\rho_r) - \log_2 N,
\label{shifted.ent}
 \end{equation}
 where $\rho_r=Tr_1...Tr_{N-1}(\ketbra{\psi}{\psi})$ is
 the single particle reduced state, { the partial trace is taken
   in the configuration space,}
 and $S(\rho)=Tr(-\rho\log_2\rho)$
 is the von Neumann entropy.
 Such a quantifier is obtained simply by noticing that
  each extremal state in 
the one particle reduced space is  respective to a unique single Slater determinant \cite{dft89}.
 {More precisely,
 the single particle reduced state of a single Slater determinant
  (as Eq.\eqref{unentangled.pure.state})
  is given by,
 \begin{equation}
 \rho_r = \frac{1}{N}\sum_{i=1}^N a^{\dagger}_{j_i} 
 \ket{vac} \bra{vac} a_{j_i}
 \end{equation} 
and its particle entanglement is null, $E_p(\ketbra{\psi}{\psi}) = 0$. If 
the state cannot be described by a single Slater 
determinant, its entanglement is necessarily 
non null, $E_p(\ketbra{\psi}{\psi})>0$, and at least one 
of its particles is entangled with another one. Maximally entangled states 
have their single particle reduced states
  described by the maximally mixed state, as for example, the strong coupling
  limit phases SDW, CDW, and PS, as described in
   Eqs.\eqref{sdw.phase.stronglimit},
   \eqref{cdw.phase.stronglimit}, \eqref{ps.phase.stronglimit}, respectively,
    whose reduced state is given by,
   \begin{equation}
   \rho_r = \frac{1}{2L} \sum_{j=1}^L \sum_{\sigma=\uparrow,\downarrow}
    \ad{j,\sigma} \ket{vac}
   \bra{vac} a_{j,\sigma},
   \end{equation}
   which have maximal von Neaumann entropy, $S(\rho_r)
   = \log_2(2L)$, and consequently maximal particle entanglement, 
   $E_p(\ketbra{\psi}{\psi}) = 1$.
}

 If the Hamiltonian has certain symmetries, its ground state
 entanglement can be analytically calculated as a simple 
function of its quadratures \cite{iemini13b}. In our 
particular case, from both  $\hat{S}_z$ and translational 
symmetries in the extended Hubbard model, formally given by,
\begin{eqnarray}
&Tr(\underbrace{\ad{{i}\sigma}a_{{j}
\bar{\sigma}}}_{\sigma \neq \bar{\sigma}}\, \rho_g) =
 0, \quad \forall i,j,& \label{Szconserva}\\
&Tr(\ad{{i}\sigma}a_{{j}\sigma}\, \rho_g) =
 Tr(\ad{({i}+{\delta})\sigma}a_{({j}+{\delta})\sigma}\, \rho_g),
&\label{translational.invariance}
\end{eqnarray} 
where $a^{(\dagger)}_{j\sigma}$ is the annihilation (creation) 
fermionic operator of a particle in the $j$th site, 
with spin $\sigma$, and $\rho_g = \ketbra{g}{g}$ is the
 ground state of the Hamiltonian, we have  that its 
single particle reduced state ($\rho_{r}({i}\sigma,{j}\bar{\sigma}) =
 \frac{1}{N}\,\,Tr(\ad{{j}\bar{\sigma}}a_{{i}\sigma} \rho_g)$) 
 is disjoint in the subspaces with distinct spin,
 $\rho_r = \rho_r^{\sigma=\uparrow} \oplus \rho_r^{\sigma=\downarrow}$,
 and each of these terms is given by a circulant matrix,
 \begin{equation}
\rho_{r}^{\sigma} = \frac{1}{N} \begin{pmatrix}
		x_0 & x_1 & \cdots & x_{L-2} & x_{L-1} \\
		x_{L-1} & x_0 & x_1 &  & x_{L-2} \\
		\vdots & x_{L-1} & x_0 & \ddots & \vdots \\
		x_2 &  & \ddots & \ddots & x_1 \\
		x_1 & x_2 & \cdots & x_{L-1} & x_0 \\
\end{pmatrix}\label{reduced.state.1D},
\end{equation}
\begin{eqnarray}
x_{\delta} &=& \mean{\ad{(j+\delta)\sigma}  a_{j\sigma}},
\end{eqnarray}
where $L$ is the lattice size. In this way the matrix is 
easily diagonalized, and its eigenvalues $\{\lambda_k^{\sigma}\}$
 are given by a Fourier transform of the quadratures,
\begin{equation}
\lambda_k^{\sigma} = \frac{1}{L}\sum_{\delta=0}^{L-1}e^{ik\delta} 
 x_{\delta} , \quad k=\left[0, \frac{2\pi}{L}, ..., (L-1)\frac{2\pi}{L}\right].
\label{eigen.mom.space}
\end{equation}
The entanglement is then directly obtained from Eq.(\ref{shifted.ent}).
 
\section{Entanglement and Quantum Phase Transitions}
\label{ent.and.qpt}

\begin{figure}
\centering
\includegraphics[scale=0.45]{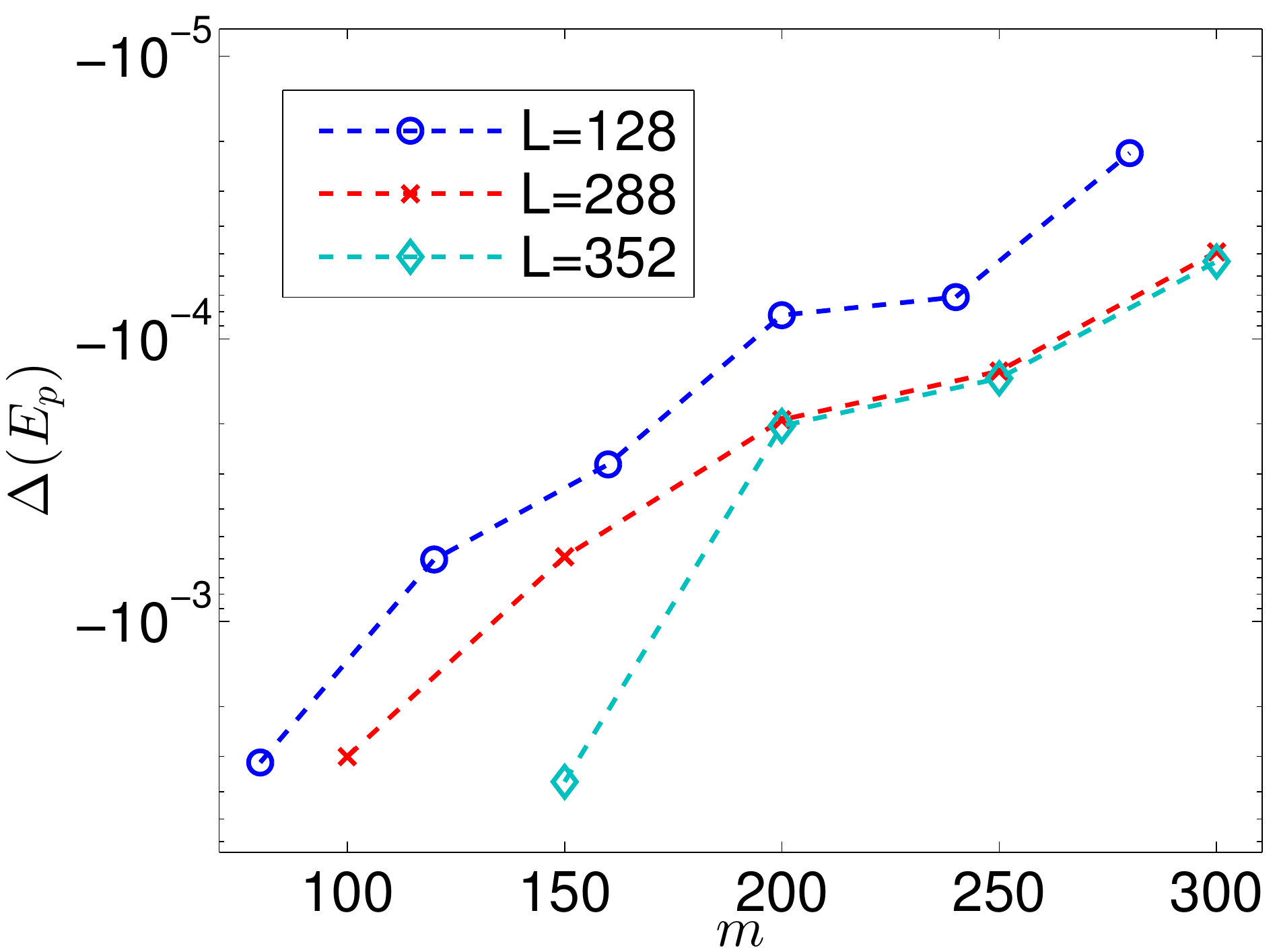}
\caption{{ (Color online) Accuracy analysis for the computation of  entanglement 
of particles using DMRG. It is shown 
the accuracy of the entanglement, $\Delta(E_p) = E_p(m) - E_p(m-50)$, as 
a function of $m$ (dimension of the renormalized matrices), at the  point $U= 4, V=2.11$, and 
using $20$ sweeps in the computation, which is enough for 
the ground state convergence. } }
\label{error.analysis}
\end{figure}

{The computation of the single particle correlations, and 
consequently the  entanglement of particles, was numerically 
performed using DMRG. 
Although DMRG is  less accurate for problems with
periodic boundary conditions (PBC) than  with
open boundary conditions (OBC), from the Physical viewpoint PBC are strongly 
preferable over OBC, as
boundary effects are eliminated and finite size extrapolations
can be performed for much smaller system sizes. 
In this work we analyze the extended Hubbard model considering 
PBC.
Our simulations were performed for 
systems up to $L = 352$ sites, always keeping a large 
enough dimension ($m$) for 
the renormalized  matrices (ranging from $m=100$ to $1000$)
and number of sweeps ($\sim 20$ sweeps), in 
order to obtain an accurate precision. 
In Fig.\ref{error.analysis},  we see  that
 $m$ ranging from $200$ to $300$  
is enough for an entanglement accuracy 
of the order of $\mathcal{O}(10^{-4})$. The accuracy for the ground state
 energy, as well as the truncation error, 
 using such parameters, are of the order of $\mathcal{O}(10^{-7})$ }

\begin{figure}
\centering
\includegraphics[scale=0.3]{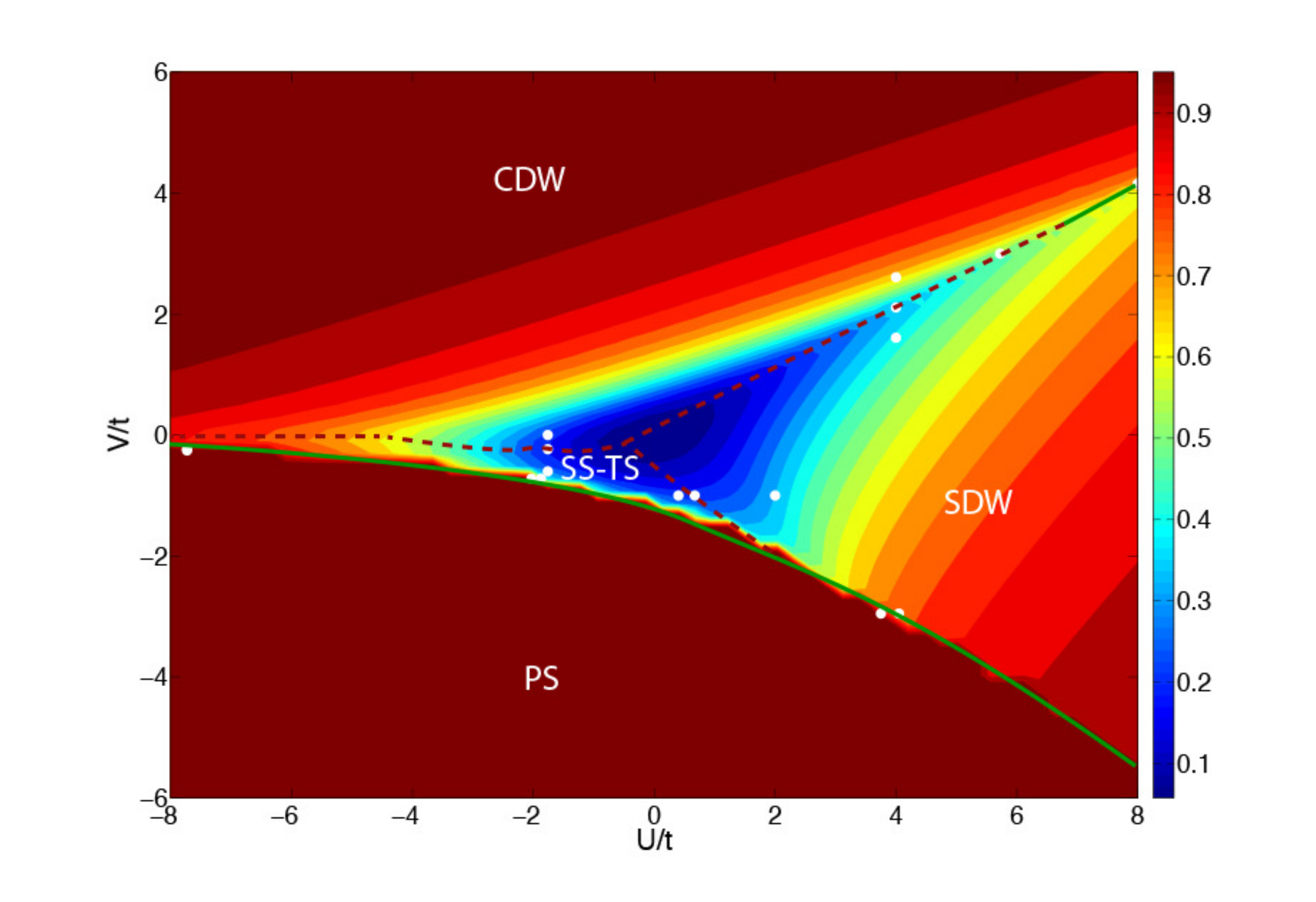}
\caption{ (Color online) {Contour map for} the entanglement of particles
 ``$E_p$''  as a function of the
 interaction terms $V/t$ and $U/t$, {in} a system with 
$L=128$ sites at half-filling. 
The entanglement behavior in the thermodynamic 
limit, $L \rightarrow \infty$, keeping fixed the filling $n= N/L = 1$, is 
qualitatively the same, with slight differences of the 
order of $\mathcal{O}(10{^{-2}})$ in its magnitude; see 
Appendix \ref{sec.finite.size.scaling} for a detailed discussion.
 The (green) continuous line denotes the  discontinuity at the 
 entanglement function, while the (red) dashed  line denotes the local minima. 
 The white dots correspond to the points where we performed a detailed finite-size scaling
  analysis (see Table I in Appendix \ref{sec.finite.size.scaling}). }
\label{phase.diagram.ent.part}
\end{figure}

Our results for 
 the entanglement of particles in the extended Hubbard
 model at half-filling are shown in Fig.\ref{phase.diagram.ent.part}, 
 to be dissected below.
 It is remarkable that such picture highlights 
the known phase diagram of the model. We first note
 that, as expected, we have a maximum of entanglement
 at the strong coupling limits ($E_{p} \rightarrow 1$), and
 as we decrease the interactions between the particles, the
 entanglement tends also to decrease, until the 
unentangled case for the non interacting Hamiltonian ($U=V=0$).
The  figure thus presents the shape of a valley around this 
point. Following then the discontinuities and the local minimum
points in the entanglement, we can easily identify
 the quantum phase transitions, 
 except for both the  subtle 
SDW-BOW transition, and the transition between 
the superconductor phases TS-SS. 
In the former case,  one needs to recall that the 
 observation of the BOW phase is by itself a hard task, 
since its gap opens exponentially
slowly, and also that there
 are evidences that such transition is of infinite order 
\cite{giamarchibook,mund09}. Therefore we believe that a 
 possible detection of such transition by the entanglement
 of particles would require
  higher precision numerical analysis as well as the study
 of larger lattice sizes. Concerning the TS-SS 
 transition, on the one hand 
 the order of the two superconducting 
 phases transition is controversial, being identified 
 as a BKT transition \cite{nakamura00} as well as a 
 second order one \cite{lin00} in the literature.  On the one hand, we 
 would  be led to strengthen the result of a BKT transition, since
  our entanglement does not detect it. 
  On the other hand,
it is  reasonable the apathy of
the entanglement of particles on distinguishing the two phases, since the
 correlations between the particles in the two superconducting 
phases have essentially the same characteristics. 
Thus it is hard to precisely conclude the reason for the
failure to  detect  such transition with our measure of entanglement.

\begin{figure}
\centering
\includegraphics[scale=0.4]{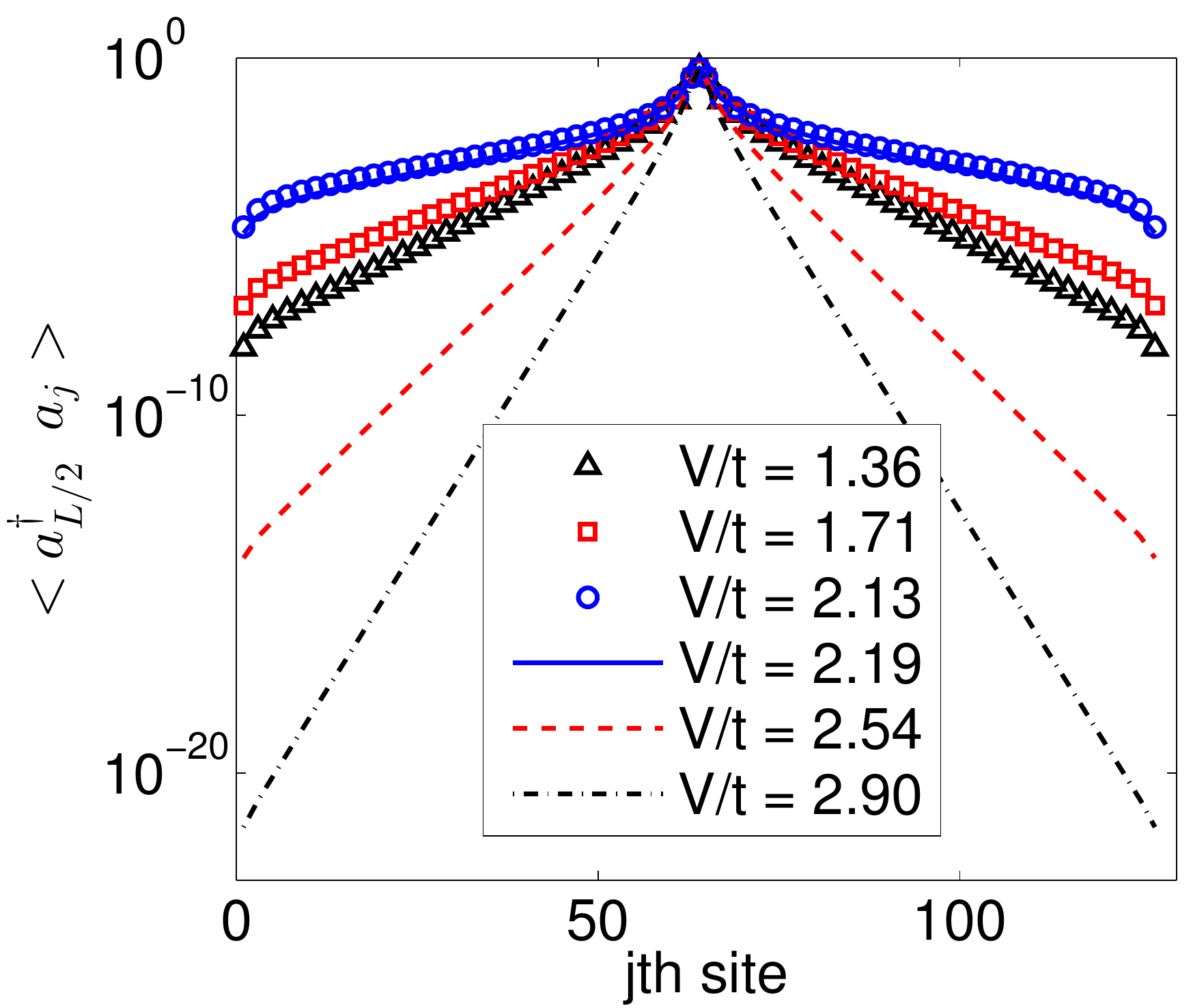}
\includegraphics[scale=0.4]{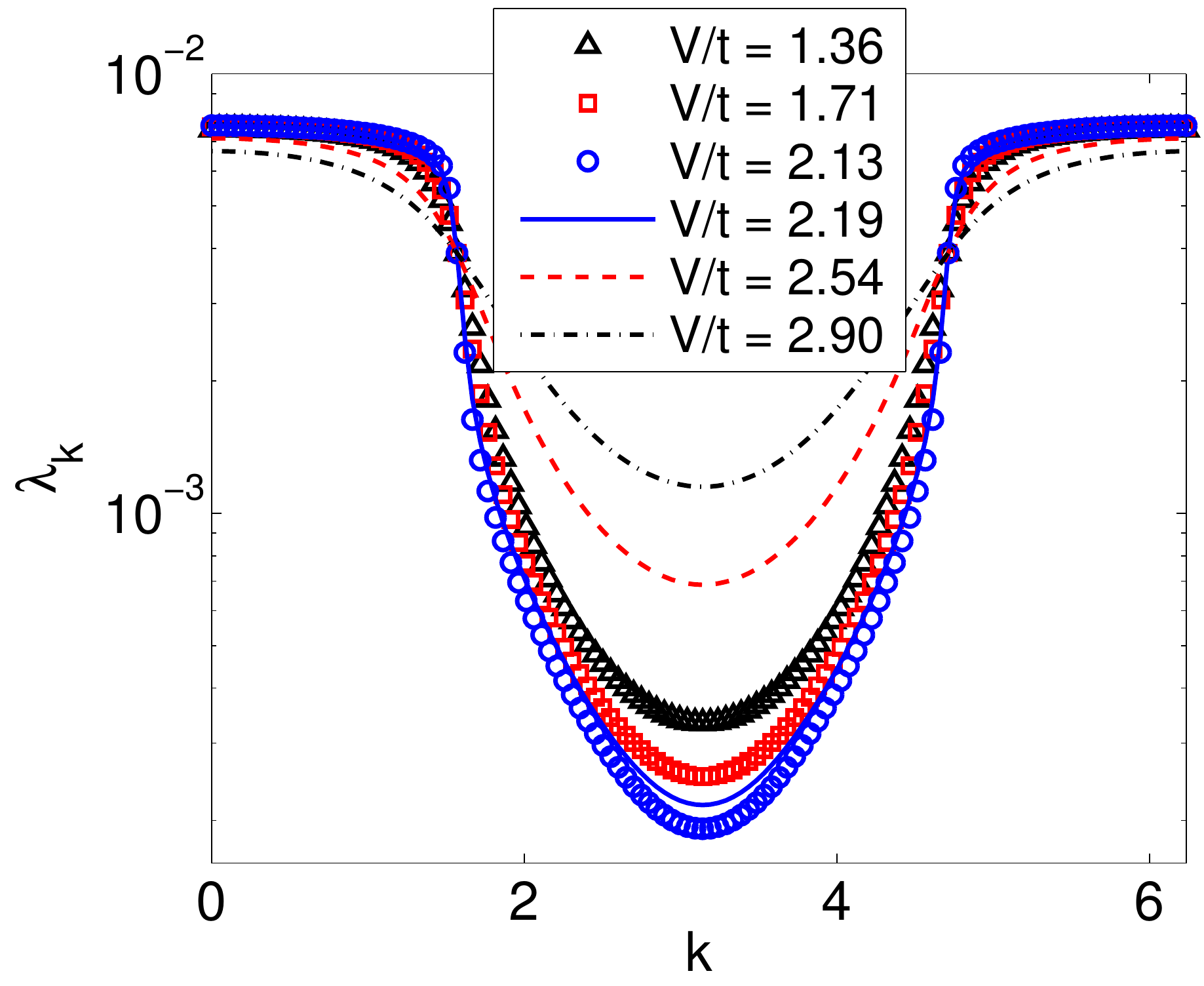}
\caption{  (Color online) \textbf{(top)} Single particle quadratures ``$\mean{\ad{L/2}a_j}$'' along the lattice sites, 
and \textbf{(bottom)}  eigenvalue distribution ``$\lambda_k$'' for the single particle reduced state in a 
fixed spin sector, as given in Eq.(\ref{eigen.mom.space}). We consider a fixed $U/t=4$.
 The vertical axis is in log-scale, in order to make clearer the visualisation.
 As we approach the BOW-CDW quantum phase transition point, at 
 {$V/t \approx2.13$},  we see that the quadratures tend to delocalise 
 along the lattice, whereas the 
 eigenvalue distribution becomes more localised.}
\label{momentum.eig}
\end{figure}

The discontinuities in the entanglement are directly 
identified with the first order quantum phase transitions,
 whereas the minimum points are identified with
  the second 
order quantum phase transitions. 
{When crossing a first order transition, the ground state
 energy presents a discontinuity and consequently also its observables. In this way, 
 the eigenvalues ``$\lambda_k$'' of the single particle reduced density matrix (Eq.\eqref{eigen.mom.space}),
  and the entanglement obtained from them, should 
  present a discontinuity.}
The occurrence of the minimum points 
 are due to the divergence of the correlation length
 when approaching the second order transitions. 
{As described} in the previous section, the eigenvalues ``$\lambda_k$''
are given in  momentum space by the Fourier transform
of the real space quadratures ``$\mean{\ad{j \sigma} a_{l \sigma} }$'' (Eq.(\ref{eigen.mom.space})).
 In this way, if we are close to the transitions, such 
real space quadratures tend to become delocalised or spread out
along the lattice, thus leading  to more localised 
eigenvalue distributions in  momentum space, and consequently
 to smaller von Neumann entropies. 
{ It is worth remarking that such behavior is the opposite of the entanglement of modes,
  where the sites are maximally entangled at the second order transitions.}
 As an example, see in Fig.\ref{momentum.eig} the eigenvalue distribution 
{for a system 
with $L=128$ sites}
  when crossing the BOW-CDW quantum phase transition.

We present now the entanglement behavior
 in some specific slices of the phase diagram {with $L=128$}, in order to 
clarify the above discussion and results. More specifically,
 we show the entanglement behavior in the
 PS-SS-CDW, PS-SS-SDW, and PS-SDW-CDW transitions.
{ Notice that our   finite-size scaling analysis showed that
  in thermodynamic limit the entanglement behavior is 
  qualitatively similar (see Appendix), 
  with a scaling inversely proportional to the lattice size,
  $E_p = a L^{-1} + b$, where $a$ and  $b$ are constants.}
  
\subsection{PS-SS-CDW}
\begin{figure}
\centering
\includegraphics[scale=0.4]{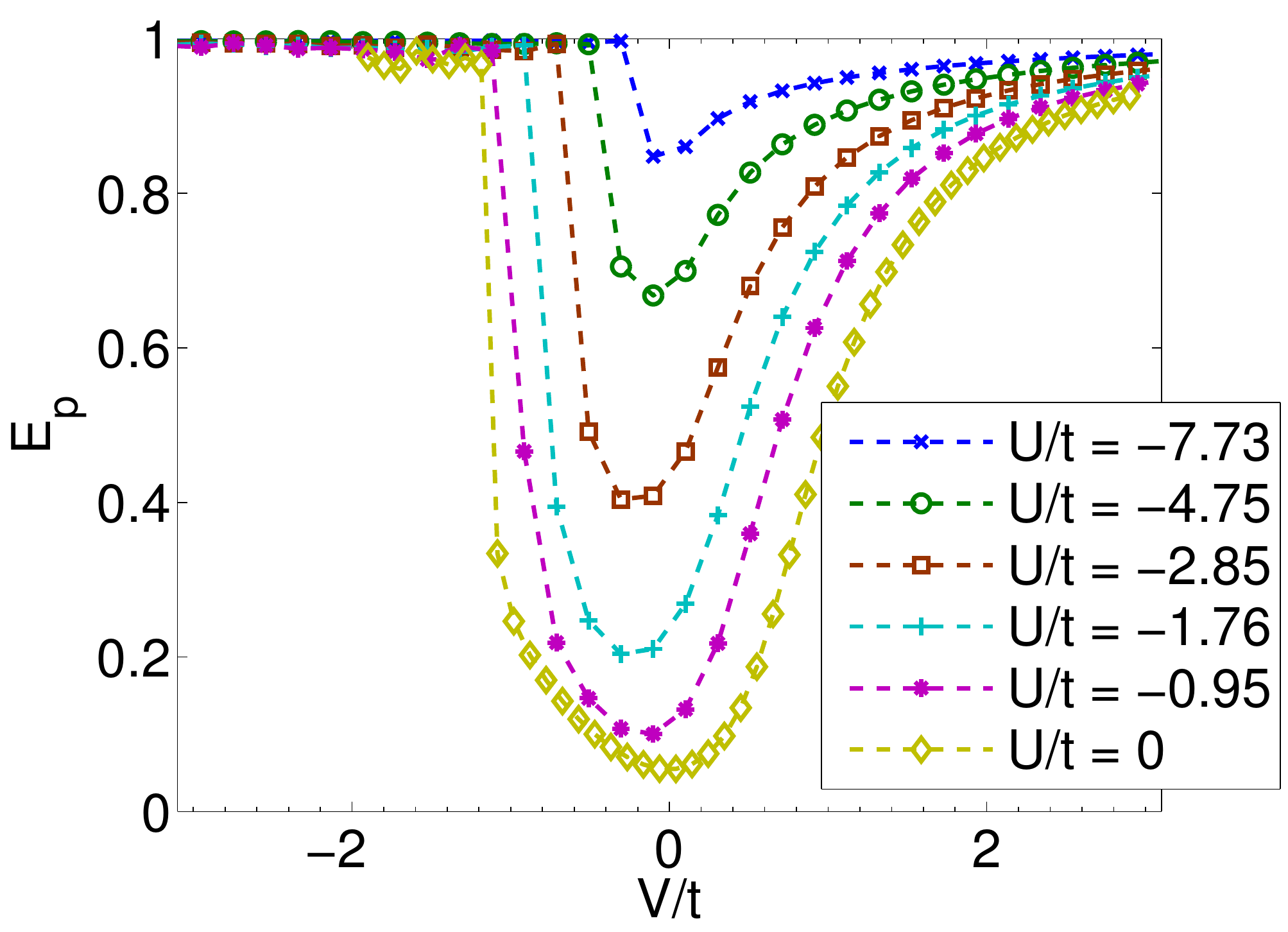}
\caption{ (Color online) Entanglement behavior across the PS-SS-CDW phases. The entanglement, for any fixed 
atractive on-site interaction ($U/t$), is characterized by a discontinuity (PS-SS transition), followed 
by a local minimun (SS-CDW transition).}
\label{slice.ps.ss.cdw}
\end{figure}

In Fig.\ref{slice.ps.ss.cdw} we see the entanglement behavior
 across the PS-SS-CDW phases. We clearly see, for any fixed
 attractive on-site interaction ($U/t<0$), a discontinuity
 in the entanglement followed by a local minimum point,
 as we increase the value of the inter-site interactions $V/t$. 
The discontinuity is related to the first order transition PS-SS, 
while the local minimum is related to the second order
 transition SS-CDW.  We see, however, that the SS-CDW transition
 is not located exactly at $V/t=0$, as expected from the phase
 diagram described in the literature, but at a value close to 
this one. We believe that this discrepancy is related to finite-size effects. 

\subsection{PS-TS-SDW}
\begin{figure}
\centering
\includegraphics[scale=0.4]{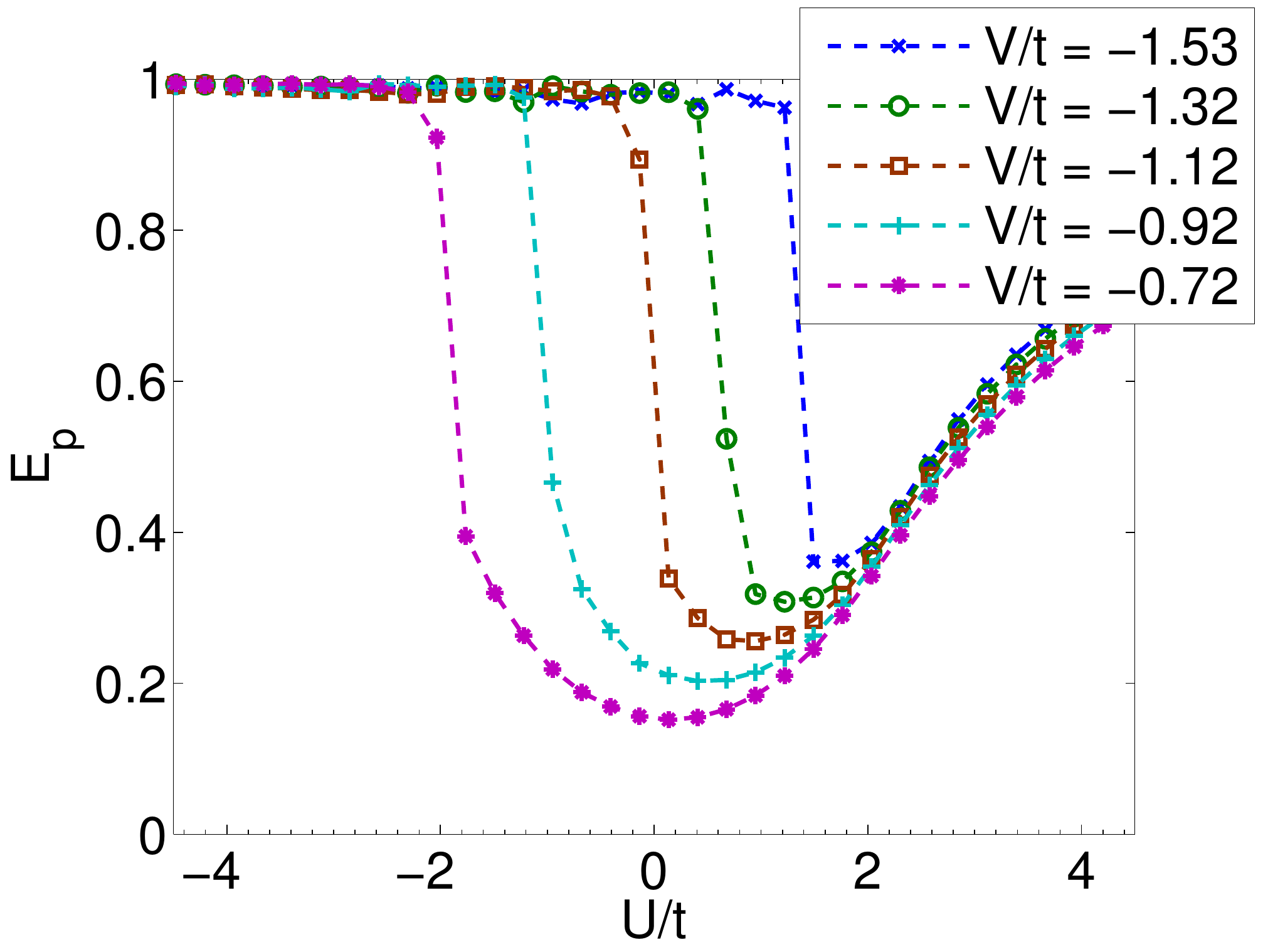}
\caption{ (Color online) Entanglement behavior across the PS-TS-SDW phases. The entanglement, for any 
fixed atractive inter-site interaction ($V/t$), is characterized by a discontinuity (PS-TS transition), 
followed by a local minimun (TS-SDW transition). For large $V/t$, the two transitions shrink at the 
same point, and there is no TS phase anymore. }
\label{slice.ps.ts.sdw}
\end{figure}

In Fig.\ref{slice.ps.ts.sdw} we see the entanglement behavior  across the PS-TS-SDW phases. We 
see again the two kinds of behavior for any fixed attractive inter-site interaction ($V/t<0$): a first 
discontinuity, related to the first order transition PS-TS, followed by a local minimum point related to 
the second order transition TS-SDW.  Note that, for large values of the attractive inter-site interaction,
 $V/t\simeq -1.5$, the discontinuity and minimum converge to  the same point, and there is no TS
  phase anymore.

\subsection{PS-SDW-(BOW)-CDW}
\begin{figure}
\centering
\includegraphics[scale=0.4]{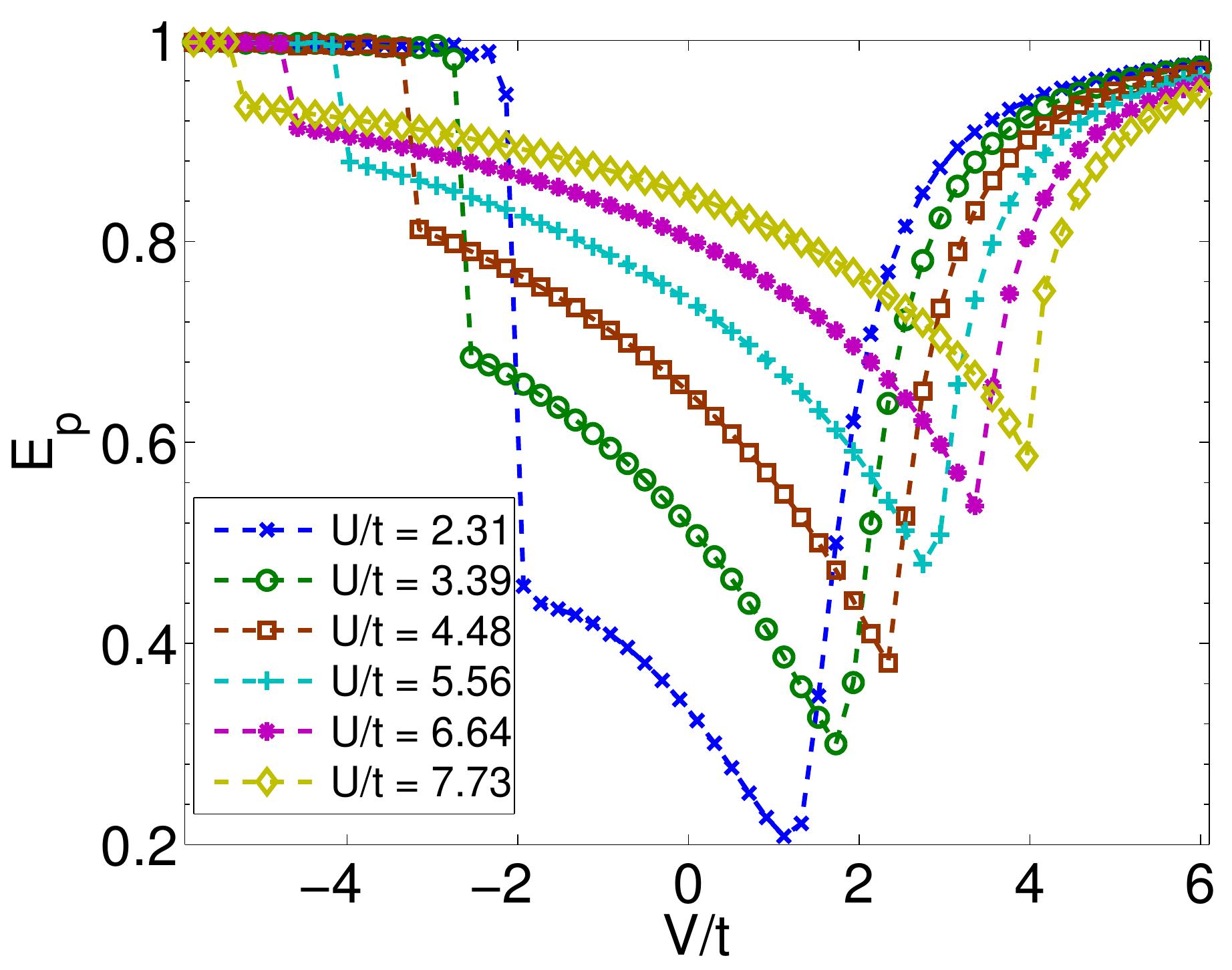}
\includegraphics[scale=0.4]{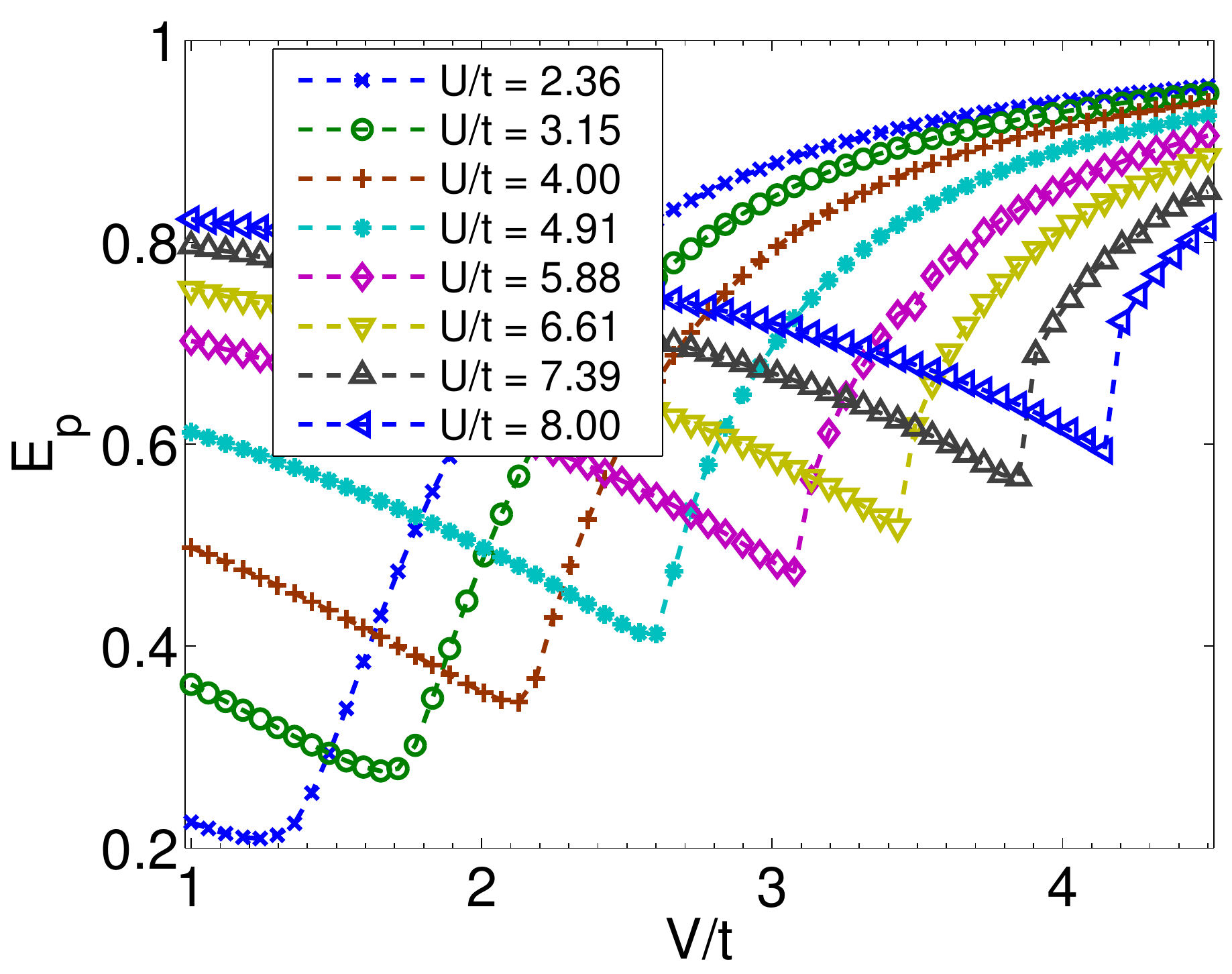}
\caption{ (Color online) Entanglement behavior across the PS-SDW-BOW-CDW phases. {The \textbf{bottom}
 panel is a magnification of the \textbf{top} panel in  the region $1\leq V/t \leq 4.5$}. The
 entanglement, for any fixed repulsive on-site interaction ($U/t$), is characterized by
  a discontinuity (PS-SDW transition), followed by: (i) a discontinuity for large $V/t$
   (SDW-CDW transition), or (ii) local minimun points for
    small $V/t$ (BOW-CDW transition). }
\label{slice.ps.sdw.bow.cdw}
\end{figure}

In Fig.\ref{slice.ps.sdw.bow.cdw} we see  the entanglement behavior across 
the PS-SDW-BOW-CDW phases. We see that, as we increase the value of
 the inter-site interactions, for any fixed repulsive on-site interactions ($U/t>0$), 
 the entanglement identifies two transitions. Firstly we see a discontinuity, related to 
 the first order transition PS-SDW, followed then by: (i)  a discontinuity, when considering 
 large $U/t$, or (ii) a local minimum point, when considering small $U/t$. Such discontinuity 
 is related to the first order SDW-CDW transition, while the minimum points are related to the 
 second order BOW-CDW transition (the SDW-BOW transition is not seen, as aforementioned).
  {We see that the transitions to the CDW phase occur at $U\approx 2V$. Performing a finite-size scaling analysis (see Appendix)
   we obtain that, for 
  $U/t=4$, the BOW-CDW transition is located at $V/t=2.11\pm0.01$, which is 
  slightly lower than the literature 
  results, namely $V/t\approx 2.16$ \cite{sengupta02,zhang04,ejima07,sandvik04}.}

\section{Conclusion}
\label{conclusion}

We studied the entanglement of indistinguishable particles
 in the extended Hubbard model at half-filling, with focus on its behavior when 
 crossing the quantum phase transitions. Our results showed that
  the entanglement either has discontinuities, or presents local minima, 
  at the critical points. We identified
 the discontinuities as  first order transitions, and the minima
 as second order transitions. In this way, we concluded that the entanglement
 of particles can ``detect''  all transitions of the known 
 diagram, except for the subtle transitions between the superconductor
 phases TS-SS, and the transition SDW-BOW.     

It is also interesting to compare our results with other entanglement measures, 
such as the  entanglement of modes, which was widely studied in several models, as
 well as in  the extended Hubbard model \cite{jian04,sa06,mund09}.  Gu \textit{et al.} \cite{jian04}
 firstly showed that the  entanglement of modes, i.e., the entanglement of a single site with
 the rest of the lattice, could {detect }three main symmetry broken phases, 
  more specifically, the CDW, SDW and PS. 
 Other phases were
 not identified due to the fact that they are associated to off-diagonal long-range order.
Further investigation  were performed analysing the block-block
  entanglement \cite{sa06,mund09}, i.e., the entanglement of a block with $l$ sites
 with the rest of the lattice ($L-l$ sites), showing that this more general measure
 could  {then detect the transition to} the superconducting phase, as well as the bond-order phase.
   The measure, however, could not {detect} the SS-TS transition, besides 
   presenting some undesirable finite-size effects in the PS phase. On the other
 hand, the entanglement of particles studied in this work showed no undesirable
   finite-size effects in the PS phase, but could not  {detect} the superconductor
  SS-TS transition either. Regarding the BOW phase, from the above discussion 
  we see that it would be worth to analyze more general measures for the
 entanglement of particles, which goes beyond single particle information.
 Some steps in this direction
 were made in \cite{ghirardi02}, where a notion
 of  entanglement of ``subgroups'' of indistinguishable particles was defined.

\acknowledgments 
We acknowledge financial support by the Brazilian agencies FAPEMIG,
CNPq, and INCT-IQ (National Institute of Science and
Technology for Quantum Information).

\appendix
\section{Finite-size scaling analysis}
\label{sec.finite.size.scaling}
\begin{figure}
\centering
\includegraphics[scale=0.42]{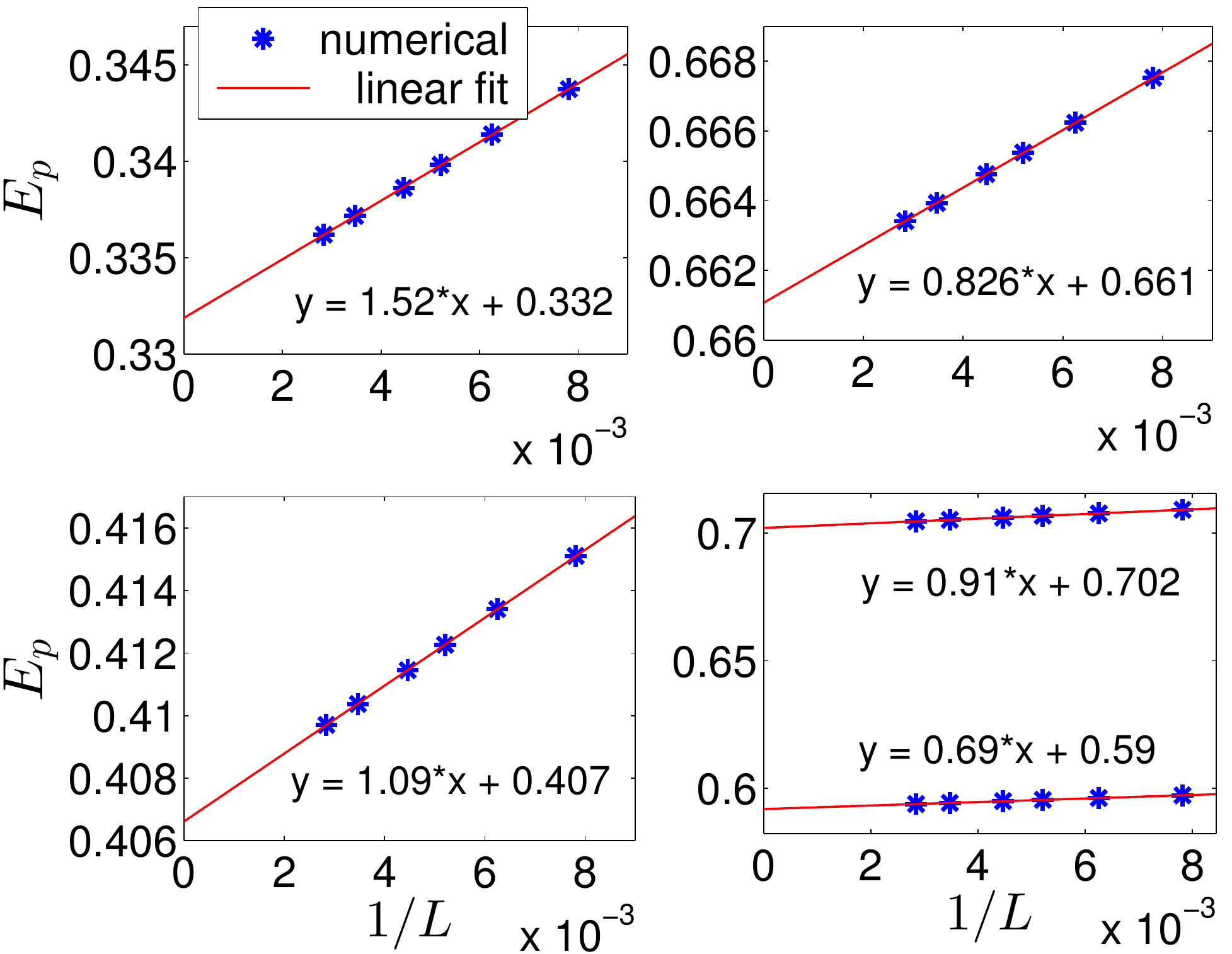}
\caption{{ (Color online) Scaling of the entanglement at 
the SDW-BOW-CDW phase transitions. For a fixed parameter $U/t=4$, we 
have: \textbf{(upper-left)} minimum entanglement critical point $(V/t)_c = 2.11$; 
\textbf{(upper-right)} $V/t= (V/t)_c + 0.5$; 
\textbf{(bottom-left)} $V/t= (V/t)_c - 0.5$ . For a fixed $U/t=8$ \textbf{(bottom-right)}, 
 there is a discontinuity in the entanglement, highlighted by the scaling at 
 $V/t = 4.147$ (bottom  curve), and $V/t = 4.153 $ (upper  curve).  } }
\label{figure.finite.scaling}
\end{figure}

{In this appendix we perform  a finite-size scaling analysis in the 
system entanglement, in order to extract  information about 
the ground state of the model. We obtained that the entanglement 
behavior is qualitatively the same for lattices larger than $L \approx 100$,
with just small differences of  the order of $\mathcal{O}(10^{-2})$ in its magnitude. 
In a general way, the entanglement scales with the inverse of the lattice size,
$E_p = a L^{-1} + b$, where $a$ and $b$ are constants. See  in Fig.\ref{figure.finite.scaling} , for 
example, the entanglement 
scaling for the SDW-BOW-CDW phase transitions. In Tab. I
we show the computed values for the scaling constants at different 
points in the phase diagram, as highlighted
 in Fig.\ref{phase.diagram.ent.part}. }

\begin{table}[h]
\centering
\label{table.finite.scaling}
\begin{tabular}{|c|c|c|c|}
\hline
$U/t (\pm 0.01)$   & $V/t (\pm 0.01)$ & $E_p(L \rightarrow \infty) (\pm 0.001)$ & $a (\pm 0.001)$ \\ \hline
4                               & 1.61         & 0.407                                 & 1.09        \\ \hline
$4^*$                          & $2.11^*$         & 0.332                                 & 1.52        \\ \hline
4                               & 2.61         & 0.661                                 & 0.826       \\ \hline
8                               & 4.147        & 0.59                                  & 0.69        \\ \hline
8                               & 4.153        & 0.702                                 & 0.91        \\ \hline
$5.723^*$                  & $3^*$            & 0.457                                 & 0.911       \\ \hline
$-1.76$                           & $-0.6$         & 0.273                                 & 2.41        \\ \hline
$-1.76^*$                  & $-0.233^*$       & 0.174                                 & 3.08        \\ \hline
$-1.76$                           & 0            & 0.211                                 & 2.51        \\ \hline
0.4                             & $-1$           & 0.211                                 & 2.11        \\ \hline
$0.67^*$			& $-1^*$           & 0.204                                 & 2.59        \\ \hline
2                               & $-1$           & 0.33                                  & 2.48        \\ \hline
$-7.73 $                          & $-2.55$        & 0.995                                 & $-$           \\ \hline
$-7.73 $                          & $-2.33$        & 0.884                                 & 0.69        \\ \hline
$-2.03 $                          & $-0.72$        & 0.978                                 & $-$           \\ \hline
$-1.875$                          & $-0.72$        & 0.403                                 & 7.59        \\ \hline
3.75                            & $-2.95$        & 0.985                                 & $-$           \\ \hline
4.05                            & $-2.95$        & 0.772                                 & 0.439       \\ \hline
\end{tabular}
\caption{{Scaling constants for the entanglement, $E_p = 
a L^{-1} + E_p(L\rightarrow \infty)$, at different points of 
the phase diagram, as highlighted in Fig.\ref{phase.diagram.ent.part}. The symbol
 ``$*$'' denotes the critical  points, and  ``$-$'' means that the entanglement 
 is constant, apart from numerical inaccuracy, for the analyzed lattices. }}
\end{table}

\end{document}